\documentclass{PoS}

\title{Constraining the Sea Quark Distributions Through W$^\pm$ Cross Section Ratio Measurements at STAR}

\ShortTitle{STAR W+/W- Ratio}

\author{M. Posik\thanks{for the STAR Collaboration}\\
        Temple University, Philadelphia, PA USA\\
        E-mail: \email{posik@temple.edu}}

\abstract{Over the past several years, parton distribution functions (PDFs) have become more precise. However there are still kinematic regions where more data are needed to help constrain global PDF extractions, such as the ratio of the sea quark distributions $\bar{d}$/$\bar{u}$ near the valence region. Furthermore, current measurements appear to suggest different high-$x$ behaviors of this ratio. The $W$ cross section ratio ($W^+$/$W^-$) is sensitive to the unpolarized quark distributions at large $Q^2$ set by the $W$ mass. Such a measurement can be used to help constrain the $\bar{d}$/$\bar{u}$ ratio. The STAR experiment at RHIC is well equipped to measure the leptonic decays of $W$ bosons, in the mid-pseudorapdity range $\left(|\eta| \leq 1 \right)$, produced in proton-proton collisions at $\sqrt{s}$ = 500/510 GeV. At these kinematics STAR is sensitive to quark distributions near $x$ of 0.16. STAR can also measure $W^+$/$W^-$ in a more forward region ranging from 1.0 $< \eta <$1.5, which extends the sea quark sensitivity to higher $x$. RHIC runs from 2011 through 2013 have collected about 350 pb$^{-1}$ of integrated luminosity, and an additional 350 pb$^{-1}$ from the 2017 run. These proceedings will present preliminary results of the 2011-2013 $W^+$/$W^-$ cross section ratio measurements. Additionally, the $W/Z$ cross section ratio, differential and total $W$ and $Z$ cross sections are presented.  
}

\FullConference{XXVII International Workshop on Deep-Inelastic Scattering and Related Subjects - DIS2019\\
		8-12 April, 2019\\
		Torino, Italy}

\begin{document}

\section{Motivation}
Parton distribution functions (PDFs) can be used to describe the internal structure of the proton~\cite{PDF}. Over the years several global analyses (CT14~\cite{CT14}, MMHT14~\cite{MMHT14}, BS15~\cite{BS15}, etc.) have used the available world data to extract PDFs from observables which are sensitive to various parton distributions.~One particular quantity that can be used to improve the PDF extractions is the $\bar{d}/\bar{u}$ ratio.~This ratio has been measured by E866~\cite{E866} and by the SeaQuest~\cite{SeaQuest} experiments. The data from the two experiments agree at $x$ < 0.2, but for $x$ > 0.25 the data precision decreases and the trends of the two data sets appear to deviate from each other. More data are needed to help further constrain the sea quark ratio.

While E866 and SeaQuest measure the $\bar{d}/\bar{u}$ ratio through Drell-Yan production, $W$ production in $pp$ collisions is also sensitive to the sea quarks. The $W^{+}$ ($W^{-}$) boson is sensitive to the $\bar{d}$ ($\bar{u}$) quark, which can be seen in equation~\ref{eq:Wdecay}.  

\begin{equation}
\label{eq:Wdecay}
u  + \bar{d} \rightarrow W^+ \rightarrow e^+ + \nu,\;\; d + \bar{u} \rightarrow W^- \rightarrow e^- + \bar{\nu}. 
\end{equation}

At leading order the $W$ cross-section ratio~\cite{Soffer94}, $\sigma_{W+}/\sigma_{W-}$, is directly proportional to the sea quark PDFs as shown in equation~\ref{eq:RW} and probes the sea quark distribution at a large $Q^2 \sim M^2_W$. 

\begin{equation}
\label{eq:RW}
\frac{\sigma_{W+}}{\sigma_{W-}} \sim \frac{\bar{d}(x_2)u(x_1) + \bar{d}(x_1)u(x_2)}{\bar{u}(x_2)d(x_1) + \bar{u}(x_1)d(x_2)}. 
\end{equation}

\section{Experiment}
The STAR experiment at RHIC~\cite{STAR} is an excellent place to measure the $W$ and $Z$ cross sections. STAR first measured them in the 2009 run that collected about 13.2 pb$^{-1}$ of data~\cite{STAR2012}. From these cross sections one can study the $W^+/W^-$ cross section ratio, $W/Z$ cross section ratio, $W$ and $Z$ differential cross sections, and the total $W$ and $Z$ cross sections. The $W$ and $Z$ cross sections were measured in $pp$ collisions at center of mass energy $\sqrt{s} = 500/510$ GeV during the 2011, 2012, 2013, and 2017 RHIC running periods. The kinematic reach of STAR allows for  complementary measurements at lower $\sqrt{s}$ compared to those performed at the LHC. Furthermore, the $W$ cross section ratio measurements also complement the  E866 and SeaQuest measurements, by accessing a larger $Q^2$ which is set by the $W$ boson mass. In the mid-rapidity region ($|\eta|\le 1$) STAR probes the $x$ range of approximately 0.1 to 0.3. There are several subdetectors used to select electrons/positrons that likely decayed from $W$ and $Z$ bosons, as well as determine their charge: the time projection chamber (TPC)~\cite{TPC}, used for particle tracking, the barrel electromagnetic calorimeter (BEMC)~\cite{BEMC} and endcap electromagnetic calorimeter (EEMC)~\cite{EEMC}, which are used to measure particle energy and for triggering. The EEMC allows the $W$ cross section ratio to be measured in the more forward direction, which will result in extending the $x$ reach of STAR to roughly 0.06 < $x$ < 0.4. The data sample used in the 2011, 2012, and 2013 analyses totals about 350 pb$^{-1}$ of data, while the 2017 analysis will add an additional 350 pb$^{-1}$ of integrated luminosity.      

\section{Results}
The $W$ and $Z$ fiducial cross sections can be measured experimentally as

\begin{equation}\label{eq:WZXsec}
  \sigma^{fid}_{W^\pm,Z} = \frac{N^{O}_{W^\pm,Z} - N^{B}_{W^\pm,Z}}{L \cdot \epsilon_{W^\pm,Z}},
\end{equation}
where $N^O$ is the number of observed boson candidates, $N^B$ is the number of background events estimated from data and Monte Carlo (MC), $L$ is the integrated luminosity, $\epsilon$ is the detection efficiency, and $W^\pm$ and $Z$ refers to the respective boson candidate. From this definition one can form differential cross sections $d\sigma^{fid}_{W^\pm,Z}/d\eta$, cross section ratios 
$\sigma^{fid}_{W^+}/\sigma^{fid}_{W^-}$, $\sigma^{fid}_{W^\pm}/\sigma^{fid}_{Z}$, and total cross sections $\sigma^{tot}_{W^\pm,Z} = \sigma^{fid}_{W^\pm,Z}/A_{W^\pm,Z}$, where $A$ is a correction factor to account for the kinematic acceptance that is not covered by the STAR detector. 

Presented here are preliminary results which are updates to a previous DIS proceedings~\cite{Posik18}. The electron and positron candidates from $W$ and $Z$ leptonic decays are selected using methodologies previously developed by STAR~\cite{STAR2012,WAL2013}, with only minimal changes to several cut values.

\begin{figure}[!h] %
\centering
\includegraphics[width=0.70\columnwidth]{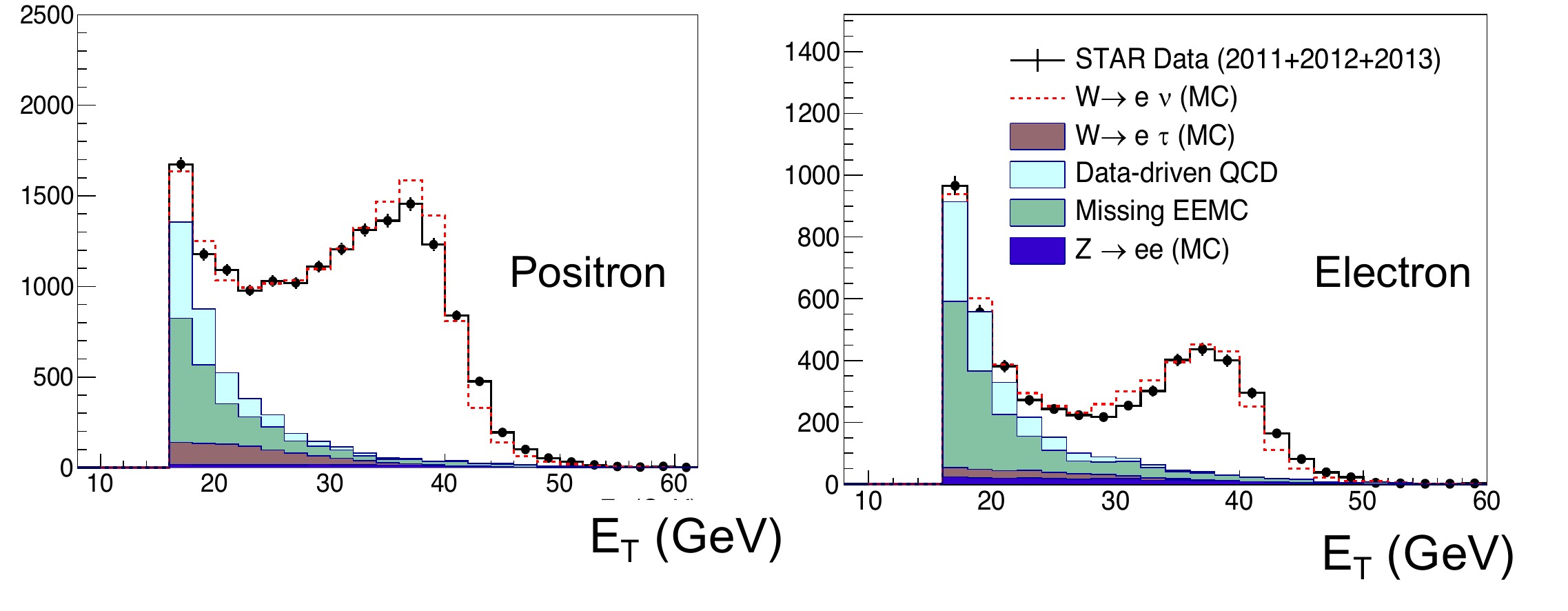}
\caption{$E_T$ distributions for $W^+$ (positrons) candidates (left panel) and $W^-$ (electrons) candidates (right panel). \label{fig:WBack}}
\end{figure}

The $W^+$ and $W^-$ background contributions measured in the BEMC, which covered $|\eta|<1.0$, for the combined 2011, 2012, and 2013 data sets are shown in Fig.~\ref{fig:WBack}. The background contributions include events from $W\rightarrow \tau + \nu$, $Z\rightarrow ee$, QCD, and missing EEMC. The QCD and missing EEMC backgrounds are estimated using the data, while the other background contributions are computed from MC. An estimate of the amount of QCD background that is present in the data is determined from the $E_T$ distribution that fails the signed-$p_T$ cut. This distribution is dominated by QCD type events. The second EEMC background is an estimate of the background caused by falsely attributing parts of jet particles to the neutrino's missing $p_T$. Also included in the figure is the $W$-decay electron signal from a MC simulation (based on Pythia 6.4.28~\cite{Pythia} and GEANT~\cite{GEANT}), and the sum of the signal and background contributions, which describes the measured $E_T$ distribution fairly well. When the final analysis cut requiring $E_T > 25$ GeV is applied, there is little background contamination remaining relative to the $W$ signal. 

\begin{figure}[!h] %
\centering
\includegraphics[width=0.4\columnwidth]{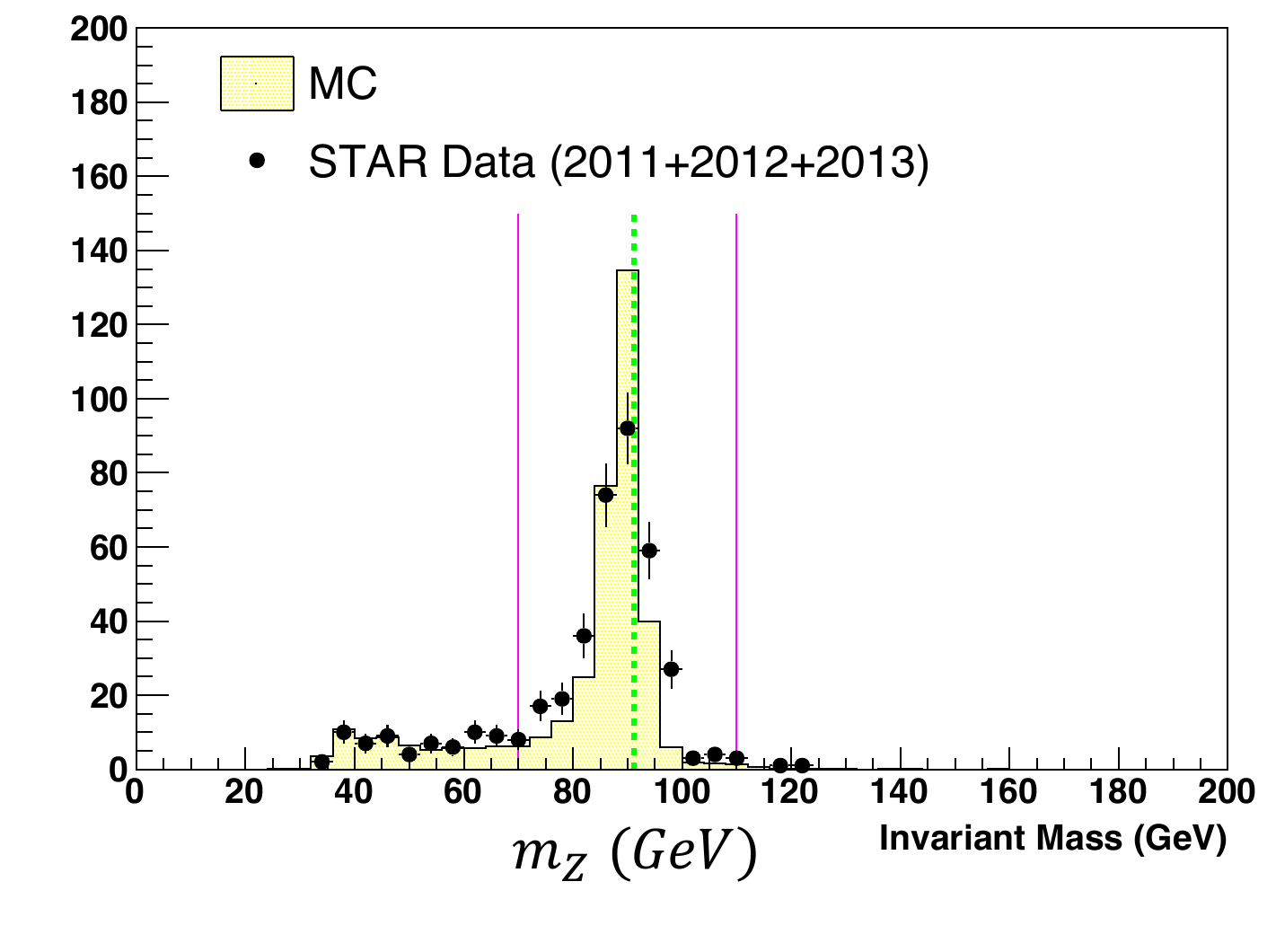}
\caption{Reconstructed invariant mass distribution of candidate $e^+e^-$ pairs from leptonic $Z$ decay. The two magenta lines show the invariant mass cut window used to select $Z$ candidates and the green dashed line marks the nominal $Z$ mass.\label{fig:ZBack}}
\end{figure}

The invariant masses of the $e^+e^-$ pairs originating from $Z$ decay can be reconstructed and is shown in Fig.~\ref{fig:ZBack}. The measured invariant mass distribution is also compared a MC simulation for $Z/\gamma^* \rightarrow e^+e^-$. Good agreement between data and MC is observed. The final $Z$ candidate selection requires the reconstructed invariant mass to be in the range of $70$ GeV  $\le m \le 110$ GeV (represented in the figure by the two magenta lines). The MC describes the measured distribution well, which suggests background contributions from any other decay channels are small. As a result no background corrections have been made to the $Z$ candidate measurements.    

\begin{figure}[!h] %
\centering
\includegraphics[width=0.8\columnwidth]{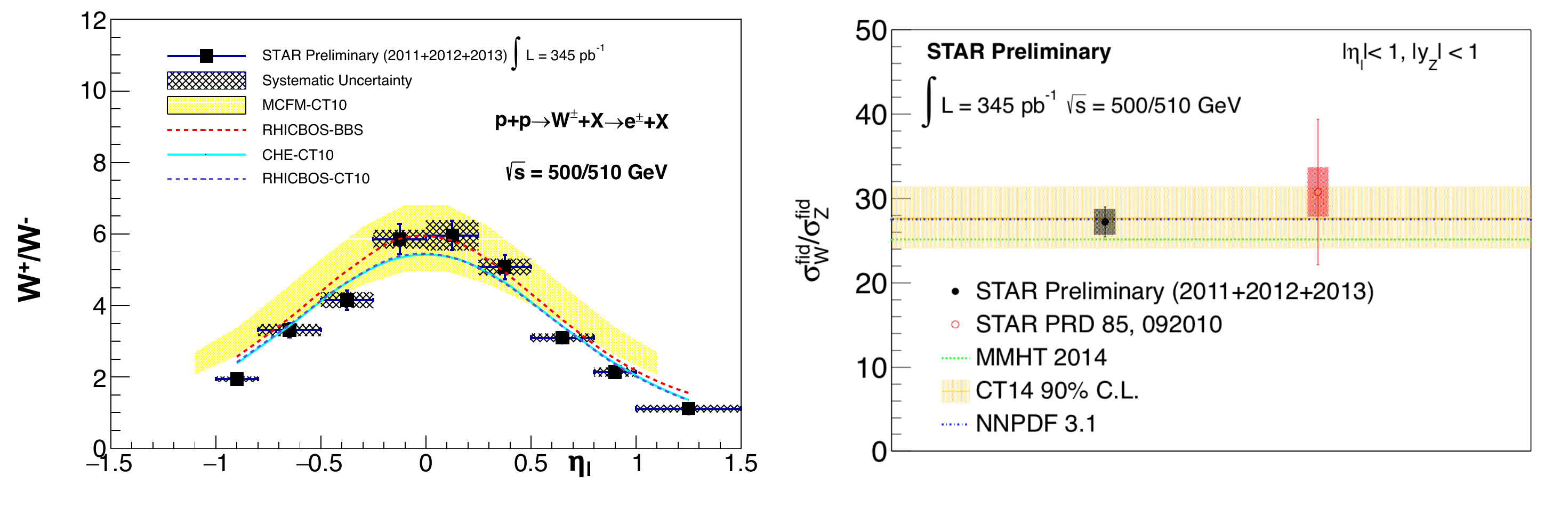}
\caption{STAR preliminary $W$ cross section ratio plotted as a function of lepton pseudorapidity (left panel). STAR preliminary $W/Z$  cross section ratio compared to that measured using the 2009 data (right panel). \label{fig:XSecRatio}}
\end{figure}

Fig.~\ref{fig:XSecRatio} (left panel) shows the $W$ cross section ratio plotted as a function of lepton pseudorapidity for the combined 2011, 2012, and 2013 data sets. The statistical uncertainties are given by the error bars, while the systematic uncertainties are represented by the shaded boxes.  The yellow band and colored curves correspond to different PDF sets~\cite{CT10nlo,BBS} and theory frame works~\cite{MCFM,RHICBOS01}.  

The $\sigma^{fid}_{W^\pm}/\sigma^{fid}_{Z}$ cross section ratio is shown in Fig.~\ref{fig:XSecRatio} (right panel), where the $W^\pm$ cross section is integrated over the range $|\eta|<1$ and the $Z$ cross section over the range $|y|<1$. The combined 2011, 2012, and 2013 result is represented by the black marker, while the 2009 data set, computed from fiducial cross sections published in Ref.~\cite{STAR2012}, is given by the red marker. The error bars represent the statistical uncertainties, while the shaded boxes represent the systematic uncertainties. The theory code FEWZ~\cite{FEWZ} was used with multiple NLO PDF sets~\cite{CT14,MMHT14,NNPDF31} to compare predictions to data. The yellow band represents the CT14 NLO PDF set at 90\% confidence level. The two STAR measurements are consistent with each other and are well described by theory. The new measurement has a smaller uncertainty than the theoretical calculation and thus will help constrain the latter.

\begin{figure}[!h] %
\centering
\includegraphics[width=0.70\columnwidth]{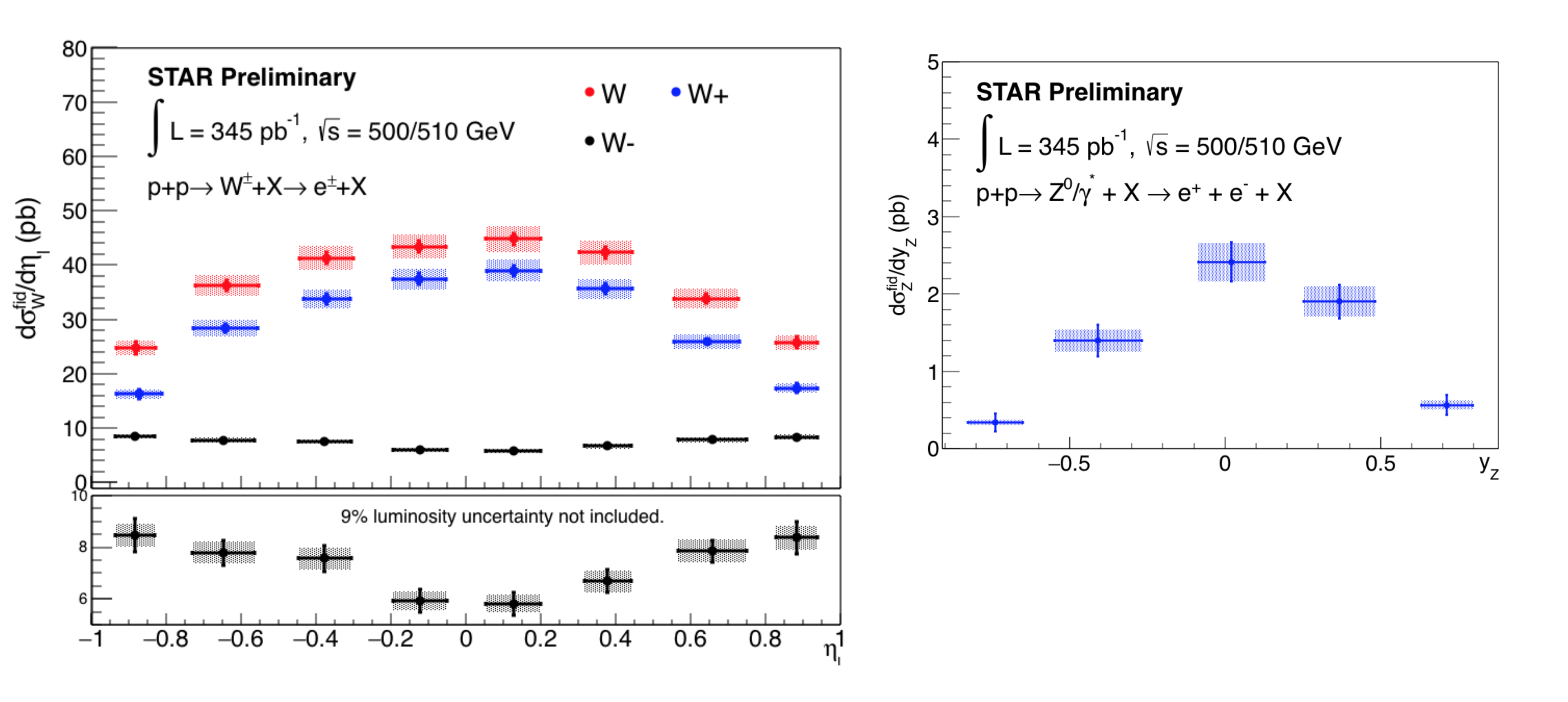}
\caption{STAR preliminary differential cross sections for $W$ (left panel) and $Z$ (right panel) bosons.\label{fig:dXsec}}
\end{figure}

Fig.~\ref{fig:dXsec} shows the differential cross section $d\sigma^{fid}/d\eta$ as a function of $\eta$ for $W^+$, $W^-$, and $W^\pm$ in the left panel, where the bottom panel zooms in on the $d\sigma^{fid}_{W^-}/d\eta$ differential cross section. The right panel shows the $Z$ differential cross section $d\sigma^{fid}_{Z}/dy$ vs. $y$. The statistical uncertainties are given by the error bars, and the systematic uncertainties are given by the shaded boxes. 

\begin{figure}[!h] %
\centering
\includegraphics[width=0.65\columnwidth]{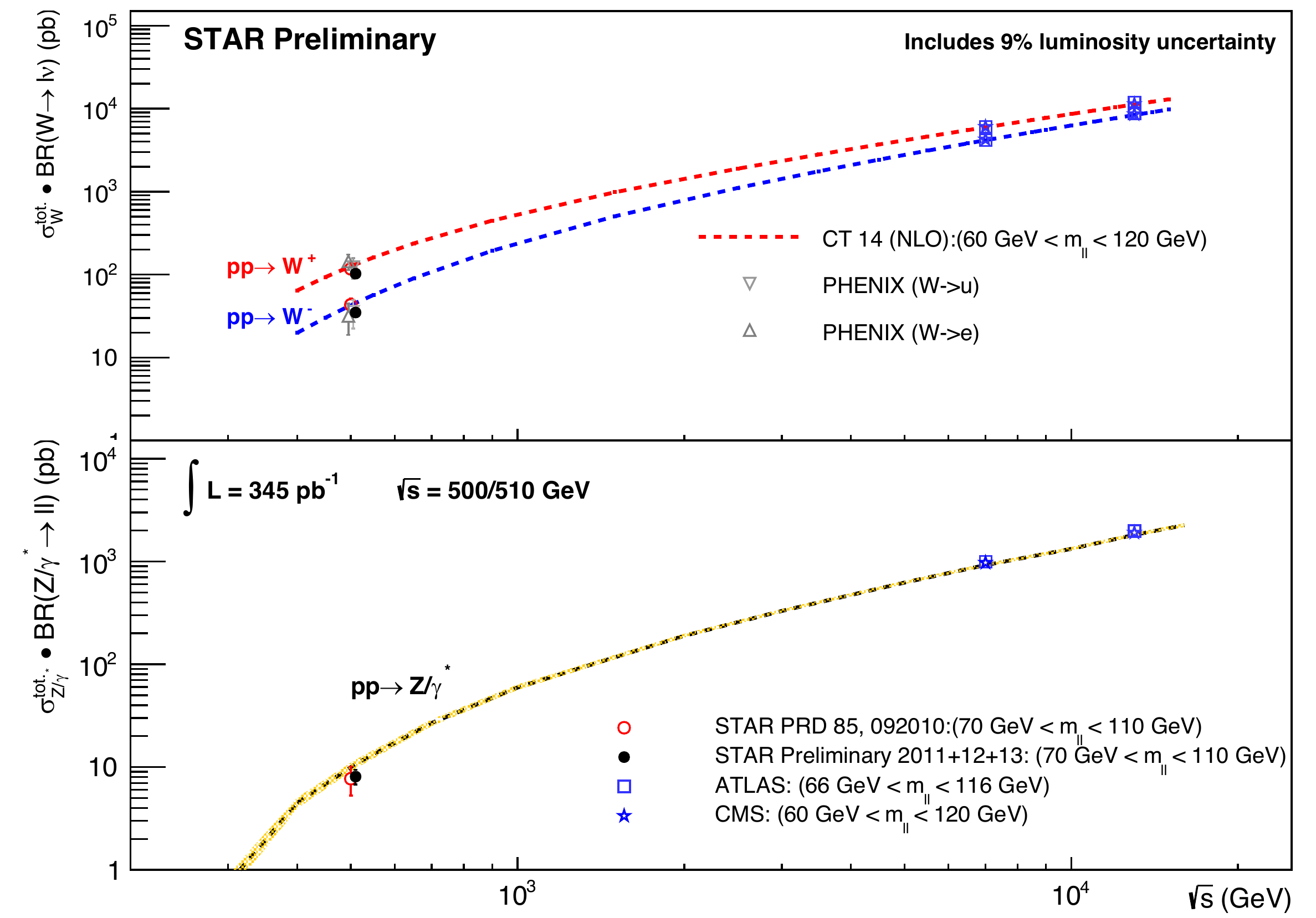}
\caption{STAR preliminary $W$ and $Z$ total cross sections vs. center of mass energy.\label{fig:Xsec}}
\end{figure}

FEWZ was used to compute the correction acceptance factor $A$ for each boson. Fig.~\ref{fig:Xsec} shows the total cross section plotted vs.~$\sqrt{s}$ and compares the STAR 2011, 2012, and 2013 combined results to previous STAR~\cite{STAR2012}, PHENIX~\cite{PHENIXXsec,PHENIXXsec2}, and LHC~\cite{ATLASZ7}-\cite{CMSXsec13} results. The cross section curves were computed using FEWZ and the CT14 NLO PDF set.  

\section{Summary}
STAR has measured the $W$ and $Z$ differential and total cross sections, along with the $W^{+}/W^{-}$ and $W/Z$ cross section ratios in $pp$ collisions at $\sqrt{s} =$ 500 GeV and 510 GeV. These measurements will provide additional high $Q^2$ data that are sensitive to the  $\bar{d}/\bar{u}$ sea quark ratio in the kinematic range of about 0.06 < $x$ < 0.4, which will help constrain the sea quark PDFs and complement the E866 and SeaQuest measurements. Furthermore, the STAR results will serve as complementary measurements to the LHC $W$ and $Z$ production measurements by providing cross section measurements at lower $\sqrt{s}$ and larger $x$. Preliminary results analyzed from the 2011, 2012, and 2013 STAR data sets, totaling about 350 pb$^{-1}$, are presented in these proceedings. The STAR 2017 run, which collided $pp$ at $\sqrt{s} = 510$ GeV, is currently under analysis and will provide an additional $\sim$ 350 pb$^{-1}$ of integrated luminosity.

\end{document}